# How can Ectrons be Accelerated by a Longitudinal Wake Field Excited in a Plasma

K.V. Ivanyan* , K.S. Badikyan

M.V. Lomonosov Moscow State University, Moscow 119991, Russia

**Abstract**

The possibility of charged particle acceleration by a longitudinal wake field excited in plasma by an electron bunch and a train of electron bunches is investigated. The exact solution of the stationary nonlinear self-consistent interaction of a monoenergetic relativistic bunch with cold plasma is obtained**.** It is shown that under certain conditions a self-acceleration of the bunch tail electrons up to high energies is possible.

## 1. Introduction

The possibility of charged particle acceleration by a longitudinal wake field excited in plasma by an electron bunch and a train of electron bunches is widely discussed in literature (see, e.g., [1-4] and references therein). One of the perspective trends in this direction is the consideration of nonlinear effects in the wake field generation. A series of works [4-8] have been done where the problem of excitation of nonlinear longitudinal stationary waves in plasma by a charged plane and an electron bunch of finite length is solved. In particular, it is shown that under certain conditions, for example, when the density of the bunch electrons $n_b$ approaches the half of the equilibrium density $n_0/2$ of plasma electrons and the condition $1 << \gamma^2 << n_b^2/n_0^2(1-2n_b/n_0)$ takes place, where $\gamma$ is the bunch relativistic factor, the longitudinal wake field intensity becomes proportional to the $\gamma$-factor, which can provide a high rate of the wake field acceleration of particles. The transformer ratio R too turns out to be proportional to the bunch $\gamma$-factor. When the inverse condition is fulfilled $\gamma^2 >> n_b^2/n_0^2(1-2n_b/n_0) >> 1$ intensity is-roportional $(1-2n_b/n_0)^{-1/2}$ (or $(1-2n_b/n_0)^{-1/4}$ , depending on the bunch length d) and can again reach large numerical values at $n_b -> n_0/2$. This case is considered in ref. [9].

In the latest years in connection with the growing interest in the nonlinear phenomena in wake fields, there appeared publications [9-11] the results of which are basically close to those of earlier works [4-11]. In ref.[11] the capture of background electrons by the wake field and the effect of thermal oscillations on nonlinear effects are also considered.

-----------------

*k.ivanyan@yandex.com

In all the works listed above [4-11] the approximation of a “given” bunch is used when the bunch parameters enter the Maxwell equations as given functions, and the effect of the bunch-excited fields on the bunch itself is not taken into account. Proper consideration of this effect can essentially change the bunch parameters, the wake field excitation conditions and may be one of the reasons for development of instability of the bunch. The dynamics of a one dimensional infinitely long bunch with homogeneous charge density in the initial moment is considered in refs. [12-14] in the framework of the perturbation theory.

In the present work the exact solution of the stationary nonlinear self-consistent interaction of a monoenergetic relativistic bunch with cold plasma is obtained when both the bunch and the plaama are described by a set of nonlinear hydrodynamics equations and Maxwell equations. It is shown, in particular, that the interaction of the bunch electrons with the bunch-excited longitudinal field leads, under certain condi- ‘Sions, to an unstable-stationary state where a part of the bunch-tail electrons have momenta much larger than the initial ones [15].

Note, that the presence in the plasma- traversing bunch of a notable amount of electrons with energies essentially higher than the initial energy of the bunch. Was still mentioned by Langmuir (see, e.g., [16] where the results of comparatlvelv recent experiments are also presented, see also [17-84] ).

## 2. Basic Equations and their Solutions

Consider the longitudinal waves ( $E_x = E_y = 0, \quad E_z \neq 0$ ) within and out of the bunch and search for stationary solutions when all the values varying in the process of interaction depend on the variable $\tilde{z} = z - V_\phi t$ , where $V_\phi$ is the phase velocity of the excited wave.

The behaviour of an electron bunch of density $n_b(\tilde{z})$ and a front moving with constant velocity $v_0 = \beta_0 c$ through a plasma of electron density $n_e(\tilde{z})$ and ions at rest is described by the following self-consistent set of equations:

$$\frac{d}{d\tilde{z}}\left(\beta\rho_e - \sqrt{1+\rho_e^2}\right) = \frac{eE_e}{mc^2}, \tag{1}$$

$$\frac{d}{d\tilde{z}}\left(\beta\rho_b - \sqrt{1+\rho_b^2}\right) = \frac{eE_b}{mc^2}, \tag{2}$$

$$\frac{d}{d\tilde{z}} n_e \left(V_e - V_\phi\right) = 0, \tag{3}$$

$$\frac{d}{d\tilde{z}} n_g \left(V_b - V_\phi\right) = 0, \tag{4}$$

$$\frac{dE_b}{d\tilde{z}} = 4\pi e\left(n_0 - n_e - n_g\right) = 0, \tag{5}$$

where $\rho_e = \rho_{ez} / mc$, $\rho_b = \rho_{bz} / mc$ are the z - components of the dimensionless momenta-of the plasma and the bunch electrons, $n_0$ is the equilibrium plasma density, $\beta = V_\phi / c$ . In case of $n_b < n_0$ and $\beta < \beta_0$, $\rho_b \le \rho_0$ is true and at boundary conditiona $E_b(d) = 0$, $\rho_e(d) = 0$ $n_e(d) = n_0$, $\rho_b(d) = \rho_0$ ($\tilde{z} = d$ - is the bunch front), from the set (1) - (5) one can obtain the nonlinear equation

$$\frac{d^2}{d\tilde{z}^2}\left(\beta\rho_b - \sqrt{1+\rho_b^2}\right) = \frac{4\pi e^2}{c^2}\left\{n_0 - \frac{n_0\beta\sqrt{1+\rho_b^2}}{\beta\sqrt{1+\rho_b^2} - \rho_e} - \frac{n_{b0}(\beta-\beta_0)\sqrt{1+\rho_b^2}}{\beta\sqrt{1+\rho_b^2} - \rho_e}\right\}. \tag{6}$$

The integration of eq.(6) with account of Eq. (6) with account of eqs.(l)- 4) and the boundary conditions, brings to the following expression for the field intensity within the bunch:

$$E_b(\tilde{z}) = \sqrt{2}\,\frac{mc\omega_p}{e}\left\{1 - \sqrt{1+\rho_b^2} + \frac{n_{b0}}{n_0}(\beta_0 - \beta)(\rho_b - \rho_0)\right\}^{1/2}. \tag{7}$$

where the variation ranges of $\rho_p$ and $\rho_e$ in the ultrarelativistic case of e $\rho_p$ , $\rho_p$ >>1 defined by

$$\begin{gathered}\rho_0 \le \rho_b \le \rho_b^0, \quad \rho_e^0 \le \rho_e \le 0, \\ \rho_e^0 = -\frac{2a\beta}{1-a^2\beta^2}, \quad \rho_b^0 = \rho_0 + \frac{2a\beta^2}{1-a^2\beta^2}\frac{1+a}{1-\beta}, \\ a = \frac{n_{b0}(\beta_0-\beta)}{n_0(1-\beta)} \Big/ \left(1 - \frac{n_{b0}(\beta_0-\beta)}{n_0(1-\beta)}\right).\end{gathered} \tag{8}$$

The second integration of eq.(6) brings to a certain implicit dependence $\tilde{z} = f(\rho_b)$ (see [15]) from which it follows that the maximum value of $\rho_b$ is achieved at the rear edge of the bunch when $d_0 = 8\frac{V_\phi\gamma_\phi^2}{\omega_p}$ at a=1 and $d_0 = \frac{\pi V_\phi}{\omega_p}$ at a << 1, and at $V_\phi \sim V_0$ is correspondingly equal to

$$\rho_b^0 = \rho_0 + 8\gamma_0^4 \quad \text{at a=1} \tag{9}$$

$$\rho_b^0 = \rho_0 + 4a\gamma_0^2 \quad \text{at a<<1} \tag{10}$$

The maximum field within the bunch $E_b^m = mc\omega_p / e$ at a= 1 ($n_{b_0} / n_0 = 1/2$) and $E_b^m = \frac{mc\omega_p}{e}\frac{n_{b_0}}{n_0}$ at a<<1 ($n_{b_0} / n_0 << 1$).

It is seen from (9) and (10) that within the bunch, in its tail, there is a possibility for electron self-acceleration. The values of $\rho_b^0$ at a = 1 are very large, but are reached at too long bunches of $d_0 = 8V_0\gamma_0^2 / \omega_p$ , while for the case with a<< 1 the bunch lengths $d_0 = \pi V_0 / \omega_p$ are not large and, e.g., the following realizable cases are possible.

At do $d_0 = 1cm$ , $n_{b_0} = 2\cdot 10^9 cm^{-3}$, $\gamma_0 = 10^2$ (50MeV), $n_0 = 2.75\cdot 10^{12} cm^{-3}$ the maximum energy gained by the tail electrons is about $\varepsilon_{\max} = mc\gamma_0(1+4a\gamma_0) = 1.3\varepsilon_0$ ,where $\varepsilon_0 = mc^2\gamma_0$ is the initial energy of the bunch electrons, and at $d_0 = 10cm$ , $n_{b_0} \sim 1.5\cdot 10^7 cm^{-3}$ , $\gamma_0 = 10^4 (5GeV)$ , $n_0 = 2.9\cdot 10^{10} cm^{-3}$ , $\varepsilon_{\max} = 21\varepsilon_0$ .

At the same time, for our considered wide bunches (the transverse dimensions god,& are the wake field lengths), the number of accelerated bunches can make l-10% of the total number of the bunch particles. In the real case of the finite transverse dimensions of the bunch it is difficult to solve the equations, but, as the preliminary calculations show, the calculated effect is conserved at least in a narrow region around the bunch axis which, however, will decrease the number of accelerated particles because of lessening of the volume where these particles are contained.

The wake field behind the bunch is found by integration of (6) at $n_{b_0} = 0$ , with account of continuity of the field E and the momentum $\rho_e$ at the rear edge of the bunch $\tilde{z} = 0$ . Since $E_b(\rho_e^0) = 0$ and $\rho_e(\tilde{z} = 0) = \rho_e^0$ , where i $\rho_e^0$ s the lower limit of in (8), the integration results in

$$E(\tilde{z} \le 0) = \sqrt{2}\,\frac{mc\omega_p}{e}\left\{\sqrt{1+\rho_b^{02}} - \sqrt{1+\rho_e^2}\right\}^{1/2}. \qquad (11)$$

From the corresponding dependence of $\tilde{z} = f(\rho_e)$ $(\tilde{z} = 0)$ [15] obtained by a twice integration of (6) at $n_{b_0} = 0$ it follows that at the bunch lengths $d_0 = 8\frac{V_\phi \gamma_\phi^2}{\omega_p}$ and $d_0 = \frac{\pi V_\phi}{\omega_p}$ the maximum intensities of the wake field correspondingly are

$$E^m = 2\frac{mc\omega_p}{e}\gamma_0 \quad \text{at a} = 1 \qquad (12)$$

$$E^m = 2\frac{mc\omega_p}{e}a \quad \text{at} \quad \text{a} << 1 \qquad (13)$$

The latter expression (13) coincides with the value of the field in a linear ap roximation. While the value of the field (12), e.g., at $n_0 = 10^{12} - 10^{13} cm^{-3}$ and $n_{b_0} = n_0 / 2$ and at $\gamma \sim 10^2$(50MeV) is $E^{m} = 4 \cdot 10^{9} V/m$ with a maximum rate of accumulation in it of $E^{m} = 4GeV/m$ at a bunch length of $d_0 = 35cm$.